# Principal modes of multimode fibers resisting fiber bending

**Jiawei Xu and Xiaosheng Xiao***
State Key Laboratory of Information Photonics and Optical Communications, School of Electronic Engineering, Beijing University of Posts and Telecommunications, Beijing 100876, China.
*Corresponding author (Email: xsxiao@bupt.edu.cn)

***ABSTRACT*** Multimode fibers (MMFs) have found wide application across various fields, such as optical communications, mode-locked lasers, and endoscopy. However, the practical use of MMFs is limited by the challenges posed by fiber bending, which leads to mode coupling. In this study, we present evidence that MMFs possess principal modes, named curved principal modes, that can resist significant bending. These curved principal modes are identified by extending the Wigner-Smith operator to curved MMFs, and are demonstrated for arbitrary bending by numerical simulations. These findings have substantial implications for mode-divide-multiplexed optical fiber communications, MMF-based endoscopy, and other related applications.

## I. INTRODUCTION

Recently, there has been renewed interest in multimode fibers (MMFs) in various fields due to their larger mode area, especially the additional spatial degrees of freedom, compared to single-mode fibers. The technique of mode-divide-multiplexing based on MMFs has been adopted to significantly increase the transmission capacity of optical communications [1]. In the fields of nonlinear fiber optics [2] and mode-locked lasers [3,4], MMFs are used as multimode nonlinear or gain media, supporting higher-dimensional nonlinear dynamics and better output performance compared to conventional systems composed of single-mode fibers. Medical endoscopes utilizing MMFs have also been proposed to reduce the diameter of conventional single-mode fiber bundle endoscopes [5].

MMFs have multiple eigenmodes , and for ideal straight, weak-guided MMFs, the eigenmodes of linearly polarized (LP) modes are widely adopted in classic fiber optics [6]. However, when propagating in a curved MMF, these LP modes couple with each other [7]. The strength of mode coupling depends on the degree of curvature, thus changes in the bending states of MMFs will modify the light transmission. For typical MMFs, mode coupling is particularly sensitive to the bending state when the bending radius is on the order of centimeters. In the practical use of MMFs, the bending of MMFs are usually unavoidable, which significantly limits their applications.

Recently, there has been attention towards using the Winger-Smith operator to search for eigenmodes in disordered media or MMFs [8-10]. For lossless optical systems, the Winger-Smith operator $Q$ can be constructed by the scattering matrix $S$ with $Q = -iS^{-1}\partial_{\omega}S$, where $\omega$ represents the optical frequency in the early literature and later could represent other concerned parameters. The eigenmodes of MMFs can be obtained by calculating the eigenstates of $Q$, which are referred to as principal modes. It has been found that the principal modes exhibit insensitivity to small variations of frequency [8]. The generalized Winger-Smith (GWS) operator was proposed in disordered media [9], where $\omega$ represents the displacement of the scattering landscape from its initial position in the $y$-direction. GWS was used to find deformation principal modes, which are insensitive to strong deformations in the transverse direction of MMFs [10]. Herein, we extend the concept of principal mode to MMFs for resisting fiber bending.

In this paper, we demonstrate the existence of eigenmodes in MMFs, termed curved principal modes, which exhibit resistance to significant fiber bending as well as to changes in bending conditions. These curved principal modes are obtained by employing the GWS operator with the parameter $\omega$ representing the curvature radius of the fibers. Through numerical simulations, we discover that these curved principal modes are present not only in circularly bent fibers with a single radius but also in fibers with arbitrarily bends. These findings will benefit various applications such as MMF-based endoscopy, mode-divide-multiplexed optical communications, and other related applications.

## II. METHOD

The GWS operator is constructed using the scattering matrix, which is the transmission matrix (in the basis of LP modes) of curved fibers in our case. This transmission matrix $H$ reflects the propagation of LP modes over a segment of MMF, and it is a non-diagonal matrix if there is coupling among the modes due to fiber bending, etc. To obtain the curved principle modes that exhibit minimal sensitivity to bending effects, we define the GWS operator for curved MMFs as follows:

$$Q_{\rho} = -\frac{i}{2}\left[H^{-1}\cdot\partial_{\rho}H - \left(H^{-1}\cdot\partial_{\rho}H\right)^{\dagger}\right] \tag{1}$$

where $\rho$ is the curvature radius of the fiber. The eigenstates $v_j$ of $Q_{\rho}$ are the curved principal modes of interest.

Herein, the transmission matrix $H$ of curved MMFs is obtained through numerically calculation, using the theory of bent MMFs [11]. $H$ is calculated as

$$H = e^{iBL} \tag{2}$$

where $L$ is the transmission length of the fiber. $B$ is defined as:

$$B_{nm} = \beta_n \delta_{nm} - \frac{n_{core} k_0}{\rho} \langle \psi_n | \xi x | \psi_m \rangle \tag{3}$$

where $\beta_n$, $\delta_{nm}$, $n_{core}$ and $k_0$ respectively represent the propagation constant of mode $n$, the Kronecker delta function, the refractive index of the core of the fiber, and the wavenumber of light in vacuum. The overlap factor $\langle \psi_n | \xi x | \psi_m \rangle$ is given by

$$\langle \psi_n | \xi x | \psi_m \rangle = \iint \psi_m^* (x, y) \psi_n (x, y) \xi (x, y) x dx dy \tag{4}$$

where $\xi$ represents the degree of compression effect caused by fiber bending, $\psi_n$ represents the mode profile of the straight fiber, and $x$ and $y$ denote the transverse coordinates of the fiber. In our case, the fiber is curved in the direction of the $x$-axis.

By using Eq. (1) and applying the following approximation to calculate the derivative of the curvature radius

$$\partial_\rho H \approx \frac{H(\rho + \delta\rho) - H(\rho - \delta\rho)}{2\delta\rho} \tag{5}$$

we can obtain the GWS operator $Q_\rho$. The eigenvectors of the matrix $Q_\rho$ correspond to the curved principle modes, in the basis of the LP modes.

## III. RESULTS

Firstly, we show the effect of bending on light transmission in MMFs. By gradually decreasing the curvature radius of the MMF, one can observe a transition in the transmission matrix $H$ from a diagonal matrix (corresponding to an ideal straight fiber with $\rho \to \infty$) to a non-diagonal, random matrix, which results in the output of a speckle pattern.

To quantify the change of the light transmission through an MMF in straight and curved states, we adopt a parameter, the fidelity $F$, which measures the similarity between the transmission matrix of the straight fiber and the bent fiber. The fidelity $F$ is given by [6]

$$F = Tr[| H(\rho) \cdot H(\rho_\infty)^\dagger |^2] \Big/ \sqrt{Tr[| H(\rho) |^2] Tr[| H(\rho_\infty) |^2]} \tag{6}$$

where $\rho_\infty$ represents the curvature radius of a straight fiber, and *Tr* denotes the trace of a matrix.

The transmission matrix $H(\rho=1)$ of a typical MMF with $\rho$=1 cm, as well as the transmission matrix $H(\rho_\infty)$ of the same fiber in a straight state, are shown in the insets of Fig. 1. For simplicity, only the coupling coefficients of the first six LP modes are given. The parameters of the MMF used in the simulations are provided in Table 1. Comparing the two matrices with $\rho$=1 cm and $\rho \rightarrow \infty$, one can observe a substantial coupling of LP modes in the bent MMF. The fidelity *F* between the transmission matrices of the straight and curved fiber is further depicted in Fig.1, illustrating a rapid decrease as the fiber curvature radius decreases.

The GWS operator, then the curved principle modes, was computed from a reference curvature radius of $\rho_{ref}$= 5 cm and a deviation of $\delta\rho$ = 10 μm, according to Eqs. (1)-(5). To evaluate the stability of the curved principal modes under bending conditions, we injected the principal modes, LP01 mode and random wave front, respectively, into the MMF and observed the corresponding outputs. Figure 2(a-c) presents a typical case, where the top and bottom figures depict the output beam profiles of these three types of modes for the MMF in straight and curved ($\rho$ =3 cm) states, respectively. It can be visualized from Fig. 2(a-c) that the curved principle mode exhibits a high resistance to fiber bending compared to other types of modes.

We further quantified the robustness of curved principal modes using the Pearson correlation coefficient $P_c$, as given by Eq. (7), which measures the correlation between the output intensity patterns obtained at each radius $\rho$ and the one at a reference curvature radius $\rho_{ref}$

$$P_c = \frac{\mathrm{cov}(u_{ref}, u)}{\sigma_{u_{ref}} \sigma_u} \tag{7}$$

where $u_{ref}$ represents the output vector (in the basis of LP modes) of the MMF curved with radius $\rho_{ref}$, while $u$ represents the output vector under other curvature radii. Figure 2(d) displays the Pearson correlation coefficients for $\rho$ =2 to 100 cm, illustrating that the curved principle modes can effectively resist fiber bending over a wide range of curvature radii. Most of the curved principal modes maintain superior resistance to interference compared to random input wave front and the LP01 mode. However, a few specific principal modes exhibit non-ideal performance, presumably due to their proximity to the cutoff, resulting in significant distortion.

In practical applications of MMFs, the bending radius along the fiber may vary. Therefore, we further considered arbitrary bending conditions. We injected the curved principal modes (calculated with the parameters of $\rho_{ref}$= 5 cm and $\delta\rho$ = 10 μm as mentioned above) into the MMF composed of ten segments with random curvature radii. The length of each fiber segment is $L/10$, and the curvature radii are $\rho = [4.4, 3.7, 3.8, 7.1, 7.2, 5.9, 5.5, 7.0, 2.6, 5.9]$ cm. The correlation coefficients between output patterns of the reference curved fiber ($\rho$=5 cm) and the randomly bent fiber are shown in Fig. 3(a). The figure presents all 59 curved principle modes, as well as the LP01, LP11a modes, and random wave front. It can be seen that most of the curved principle modes perform better than the LP01 mode, and all the curved principle modes are more resistant to fiber bending than the LP11a mode and random wave front.

We also verified the existence of curved principal modes for another bending condition, where the fiber is composed of ten segments with random curvature radii ($\rho$ in the range of 4 to 6 cm). By use of these curvature radii, the curved principle modes are numerically calculated for this bending condition. The stability of these principle modes was estimated by changing all the curvature radii by the same amount $\Delta\rho$. Similar to Fig. 2(d), Fig. 3(b) compares the Pearson correlation coefficients of the curved principle modes to the LP01 mode and random input wave front. The results demonstrate notable resistance to the bending effect of the curved principal modes compared with random input and LP01 mode.

## IV. CONCLUSION

We have demonstrated the existence of curved principal modes in MMFs and shown their ability to withstand significant fiber bending. The principal modes are obtained by extending the GWS operator for bent MMFs, exhibiting unchanged output patterns over a wide range of curvature radii (as low as 4 cm, depending on the MMF parameters). Furthermore, our findings indicate that these curved principle modes can even resist arbitrary fiber bending, making them highly valuable for practical applications of MMFs. These results are of considerable significance for future experimental verification and have practical implications for the design and optimization of fiber systems in various applications, ensuring reliable and efficient performance.

**Acknowledgements**

This work was partially supported by the National Natural Science Foundation of China (62375024).

## Figures

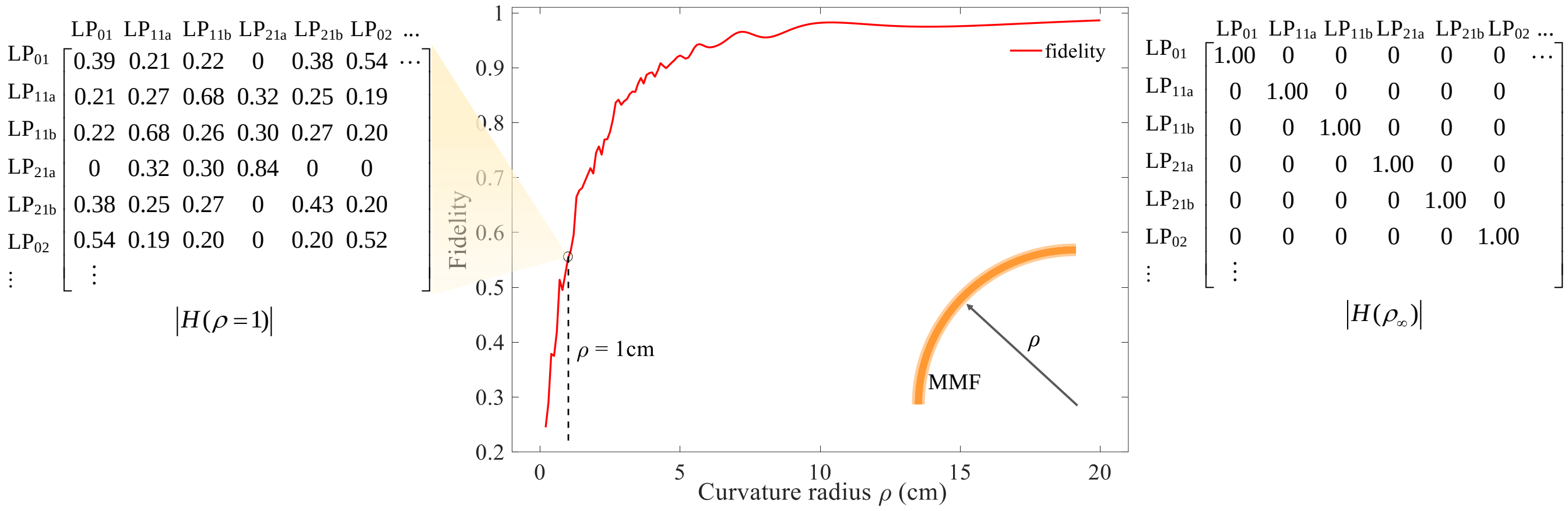

FIG. 1 Fidelity between the transmission matrices of a MMF in straight and curved states.

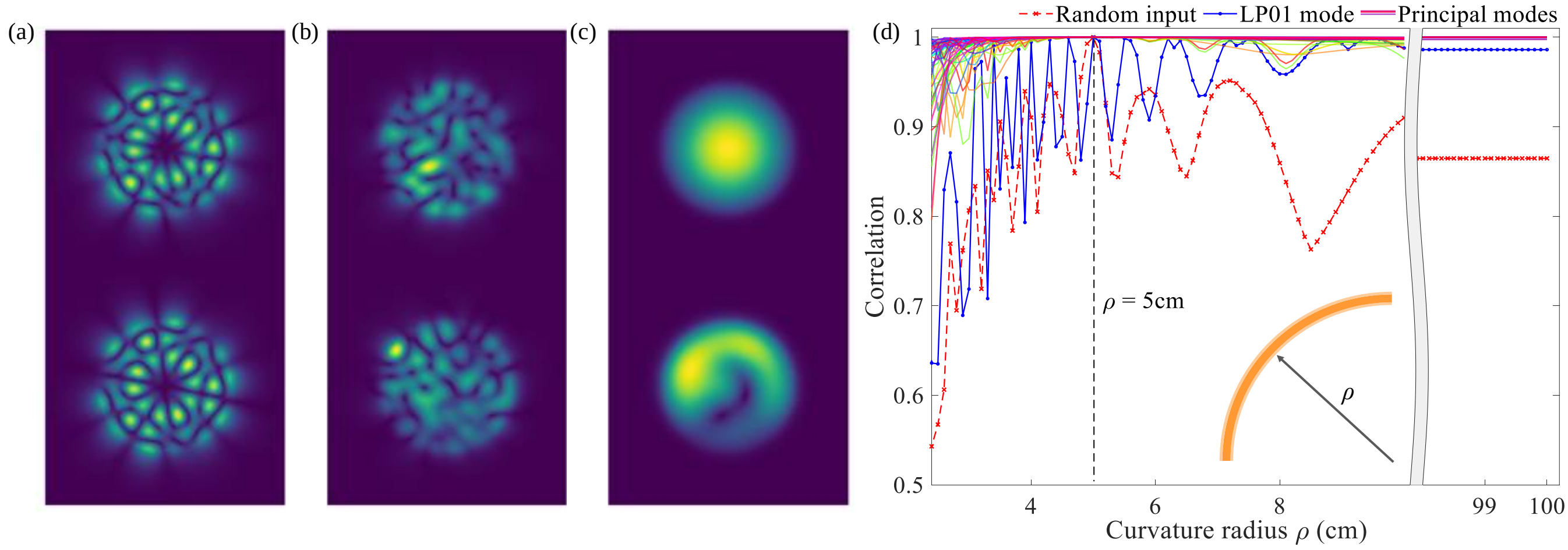

FIG.2 Comparison of curved principle modes, a random wave front and LP01 mode for a typical MMF. Output intensity profiles for the injection of (a) a curved principal mode, (b) a random wave front and (c) LP01 mode to the MMF in straight (top) and curved (3 cm radius, bottom) states, respectively. (d) Correlations between output patterns of the MMF curved with $\rho_{ref}$ = 5 cm and other radius $\rho$ for random wave front input, LP01 mode and principal modes.

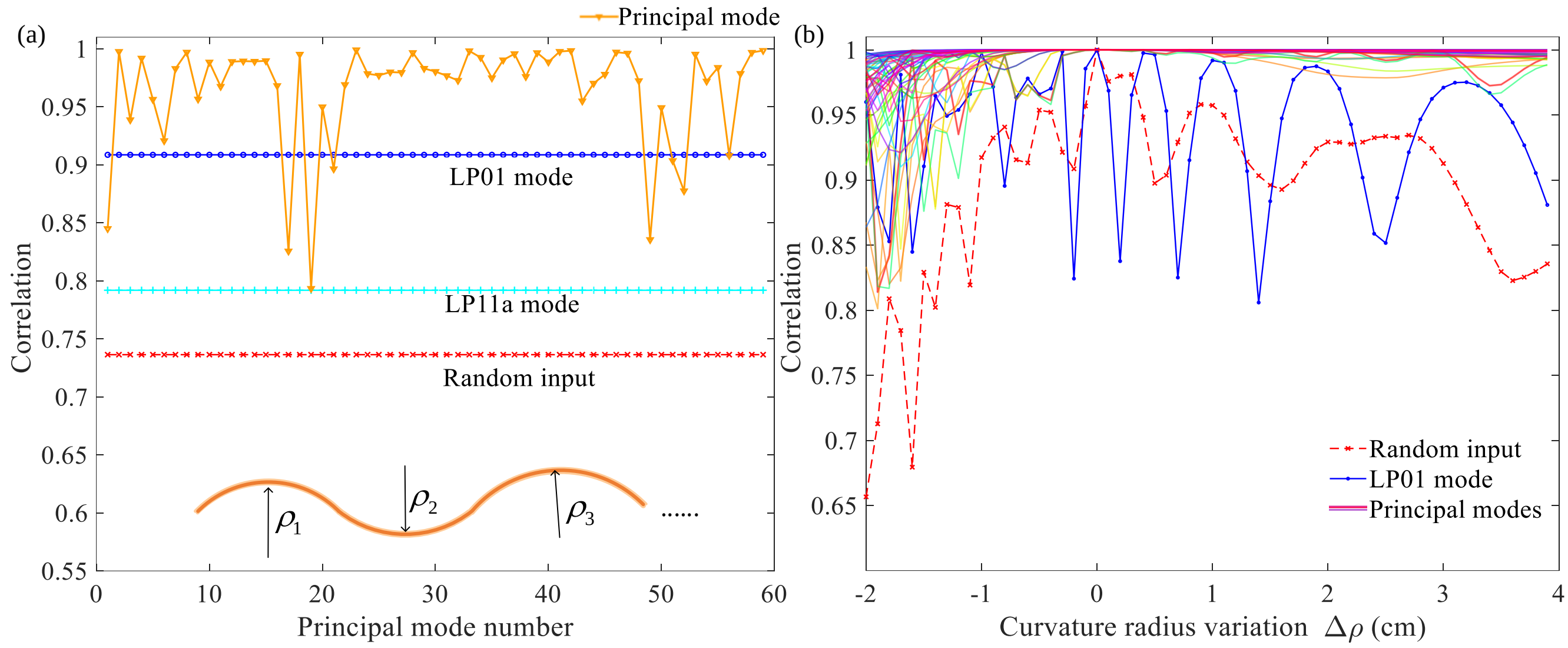

FIG.3 Correlation coefficients of output patterns for randomly bent MMFs, with the inputs are curved modes, LP modes, and random wave front, respectively. (a) Correlations between output patterns of MMF bending with a single radius ($\rho_{ref}$= 5 cm) vs randomly bending (composed of ten segments with random curvature radii, as illustrated by the inset). (b) Correlations between output patterns of the MMF with different bending conditions. Reference case: the MMF composed of ten segments with different curvature radii $\rho_i$ (where $i$=1 to 10), Correlation calculation case: the MMF bending with $\rho_i+\Delta\rho$ (where $i$=1 to 10).

## Table

Table 1. Simulation parameters of the MMF used

| Parameter | Symbol (unit) | Value |
|---|---|---|
| Wavelength | $\lambda$ ($\mu m$) | 632.8 |
| Numerical aperture | $NA$ | 0.15 |
| Core refractive index | $n_{core}$ | 1.45 |
| Core diameter | $r_{core}$ ($\mu m$) | 10 |
| Length | $L$($cm$) | 10 |